\documentclass{ws-procs961x669}            % default, citations in superscript
\begin{document}
\title{Theoretical Implications of Precision Measurements}

\author{Jens Erler}

\address{Instituto de F\'isica, Universidad Nacional Aut\'onoma de M\'exico, \\
Apartado Postal 20--364, CDMX 01000, M\'exico\\
$^*$E-mail: erler@fisica.unam.mx\\
www.unam.mx}

\begin{abstract}
Results of an updated global electoweak fit to the Standard Model (SM) are presented, 
where special attention is paid to some key observables, such as the weak mixing angle, the $W$ boson mass, 
and the anomalous magnetic moment of the muon.
Implications for new physics beyond the SM are also discussed.
\end{abstract}


\bodymatter

\section{The Electroweak Fit}

As of this writing the Standard Model (SM) is almost exactly 50 years old.
In the fall of 1967 Steven Weinberg proposed a model of leptons~\cite{Weinberg:1967tq}
which was both inconsistent and incomplete due to the lack of the quark degrees of freedom,
rendering it gauge anomalous but this was unknown at the time.
It was then modified several times, and evolved into what we call the SM today,
which despite of its senior age, appears almost immortal. 

And, of course, on the 4th of July was the 5th birthday of the Higgs boson.
To celebrate it let us time travel back to the time of ICHEP~2012 where the discovery was announced.
If one took all of the available data at that time except for the LHC reconstruction data of the Higgs boson itself,
namely electroweak precision measurements and the exclusion regions mapped out at other facilities,
one found~\cite{Erler:2012uu} the probability distribution of the Higgs boson mass, $M_H$, shown in Fig.~\ref{ICHEP}(a).
Including then the direct measurements of $M_H$ at the time, one obtained the distribution in Fig.~\ref{ICHEP}(b),
showing a remarkable consistency and demonstrating the important role that the precision data played
in terms of predicting $M_H$ before its discovery and contributing to its rapid acceptance.

\def\figsubcap#1{\par\noindent\centering\footnotesize(#1)}
\begin{figure}[t]
\begin{center}
\hspace*{5pt}\vspace{5pt}
\parbox{2.1in}{\hspace*{-25pt}\includegraphics[width=200pt]{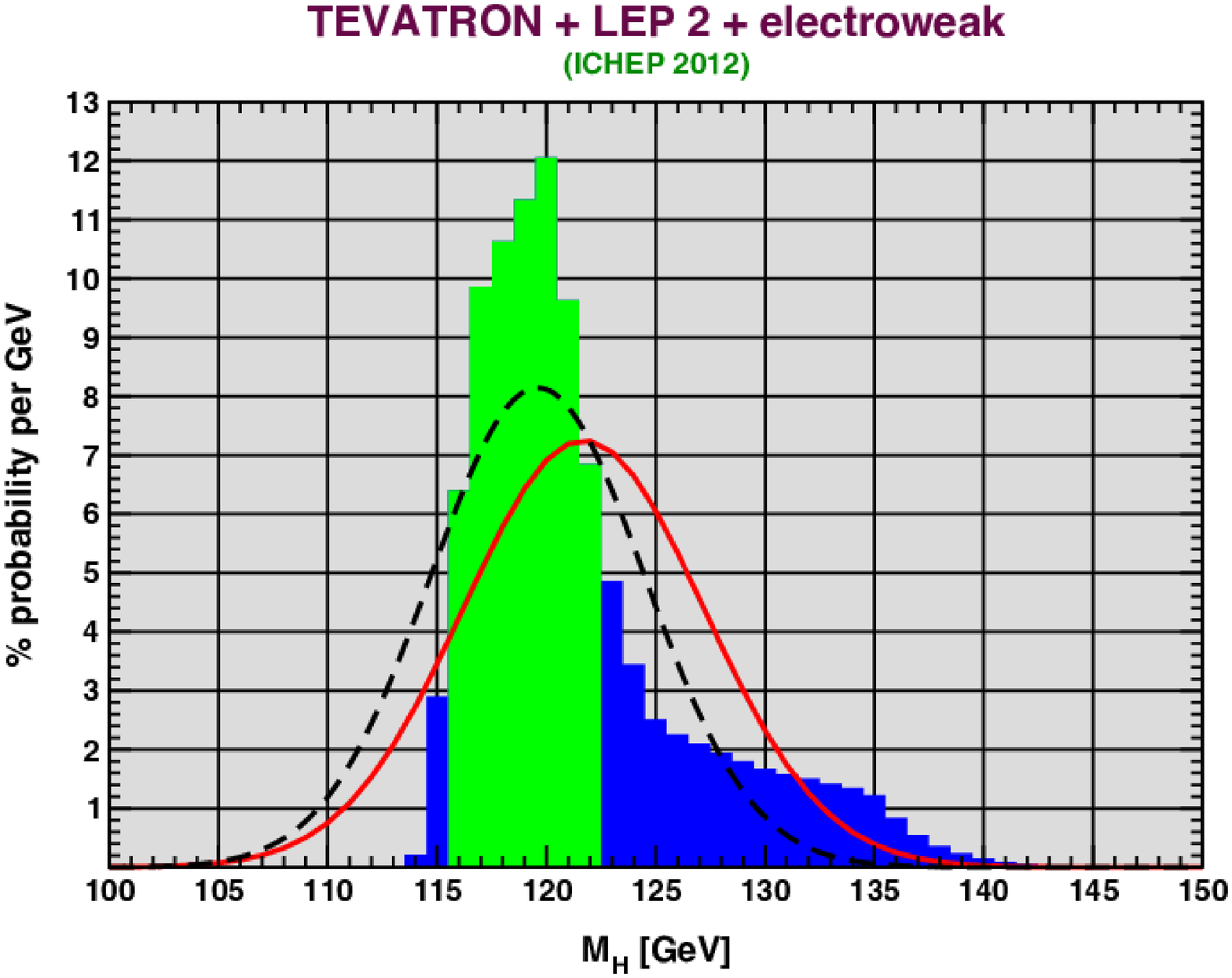}\figsubcap{a}}
\hspace*{25pt}
\parbox{2.1in}{\hspace*{-25pt}\includegraphics[width=200pt]{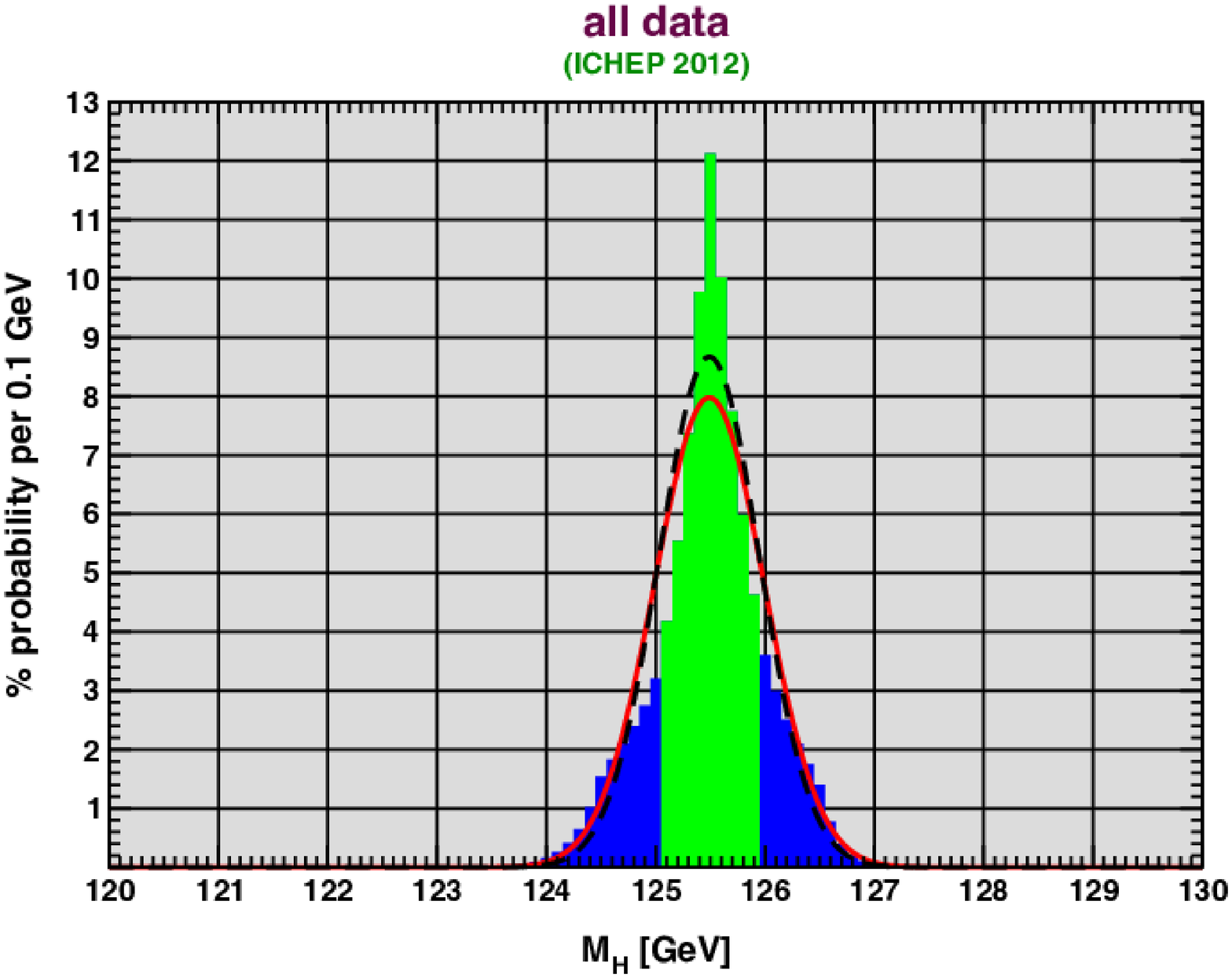}\figsubcap{b}}
\caption{Normalized probability distributions of $M_H$ where the 68\%~CL highest probability density regions are marked in green.
Also shown are two reference Gaussian: 
the dashed ones (in black) are centered around the median and have the same width as the 68\%~CL {\em central\/} 
probability intervals while the solid ones (in red) are based on mean and variance.
(a) Electroweak precision data supplemented with exclusion constraints from the Tevatron and LEP~2 as of ICHEP 2012. 
(b) All data at that time.} 
\label{ICHEP}
\end{center}
\end{figure}

This brings us to old physics implications of precision measurements, also known as the global electroweak fit. 
Some of the results presented here are from the 2016 edition of the {\em Review of Particle Physics}~\cite{Patrignani:2016xqp},
but I also include updates due to new results of this year. 
Besides the Yukawa sector, there are basically five parameters needed to fix the bosonic sector of the SM,
namely the $SU(3)_C \times SU(2)_L \times U(1)_Y$ gauge couplings and two parameters from the Higgs potential.
One combination of these is the fine structure constant, $\alpha$,
which one can take, for example, from the Rydberg constant, and doing so would leave the even more precise determination 
from the anomalous magnetic moment of the electron as a derived quantity and an extra SM test. 
Another combination is the Fermi constant, $G_F$,
which is extracted from the precise muon lifetime measurement~\cite{Webber:2010zf} at the PSI in Switzerland.
One obtains the $Z$ boson mass, $M_Z$, from the $Z$  line shape fit~\cite{ALEPH:2005ab} at LEP,
and nowadays $M_H$ from the LHC~\cite{PianoriLP2017} (see Sec.~\ref{secmh}).
On the other hand, the strong coupling constant, $\alpha_s$, is left as an unconstrained fit parameter and is thus an output.

\subsection{Weak probes of the strong coupling}
There are basically two opportunities to obtain $\alpha_s$ from electroweak observables.
One is again from the $Z$ line shape, namely from the total decay width and the hadronic peak cross section,
in addition to various branching ratios. 
The resulting constraint,
\begin{equation}
\alpha_s(M_Z)  = 0.1203 \pm 0.0028\ ,
\end{equation}
is the only one not limited by QCD theory.
Incidentally, another way to use the $Z$ line shape is to fit to the number of neutrinos, $N_\nu$.
In the past the result has been somewhat low compared to the SM prediction of $N_\nu = 3$, but now we find~\cite{PDG2016},
\begin{equation}
N_\nu  = 2.992 \pm 0.007\ .
\end{equation}
The $W$ boson width is also strongly dependent on $\alpha_s$, but due to its lower precision 
it rather serves as a first plus second row CKM unitarity test.

The other possibility to obtain a determination of $\alpha_s$ from a weak process is through the $\tau$ lepton lifetime and its branching ratios.
The $\tau$ mass is at a much lower energy scale where QCD is at the verge of a perturbative breakdown.
Indeed, one can express the perturbative expansion in two different ways, either strictly in fixed-order perturbation theory, 
or by re-summing certain terms to all orders which is referred to as contour-improved perturbation theory~\cite{LeDiberder:1992jjr}.
Both of these expansions work quite well, but while they are seemingly converging they do not appear to converge to the same value.
The reason for this is not at all understood and introduces an additional uncertainty.
Including higher order terms in the operator product expansion and quark-hadron duality violating corrections~\cite{Boito:2012cr,Boito:2014sta},
as well as the associated uncertainties, we find,
\begin{equation}
\alpha_s(m_\tau)  = 0.314^{+0.016}_{-0.013}\ , \qquad\qquad\qquad
\alpha_s(M_Z) = 0.1174^{+0.0019}_{-0.0017}\ .
\end{equation}

In total, the global fit which also contains many other $\alpha_s$-dependent quantities returns the result~\cite{PDG2016},
\begin{equation}
\alpha_s(M_Z) = 0.1182 \pm 0.0016\ .
\end{equation}

\begin{table}[t]
\tbl{Combinations of measurements of the top quark mass$^{\text a}$}
{\begin{tabular}{@{}ccccc@{}}
\toprule
& central value & statistical error & systematic error & total error \\
\colrule
ATLAS & 172.84 & 0.34 & 0.61	& 0.70 \\
Tevatron & 174.30 & 0.35	& 0.54 & 0.64 \\
CMS & 172.43 & 0.13 & 0.46 & 0.48 \\
\colrule
grand average & 172.97 & 0.13 & 0.38 &0.41 \\
\botrule
\end{tabular}}
\begin{tabnote}
$^{\text a}$ All values in GeV. \\
\end{tabnote}
\label{mt}
\end{table}

\subsection{Top quark mass}
Let us now turn to measurements~\cite{YamazakiLP2017} of the top quark pole mass, $m_t$.
ATLAS~\cite{Nisius:2017ppa}, CMS~\cite{Spannagel:2016cqt}, 
as well as the Tevatron Electroweak Working Group~\cite{TevatronElectroweakWorkingGroup:2016lid} each produced a combination 
(shown in Table~\ref{mt}) of their various determinations of $m_t$, based on different decay modes and analysis details.
Assuming a common theoretical uncertainty for all three combinations equal to the smallest one,
which is the quoted 0.29~GeV uncertainty from ATLAS, it is not difficult to perform the grand average,
\begin{equation}
m_t  = 172.97 \pm 0.28_{\rm uncorr.} \pm 0.29_{\rm corr.} \pm 0.50_{\rm QCD}~{\rm GeV}\ ,
\label{mtpole}
\end{equation}
where the first and second error are the uncorrelated and correlated components, respectively.
However, to split the total error of this average up into statistical and systematic components as done in Table~\ref{mt}
is less straightforward, and for the issues involved I refer the interested reader to Ref.~\citenum{Erler:2015nsa}.  
The last error in Eq.~(\ref{mtpole}) from QCD includes the uncertainty from the ambiguity 
which top quark mass definition is actually measured at a hadron collider, and 
given that it is presumably close to the pole mass one also needs to include an extra error from the conversion~\cite{Marquard:2016dcn}
from the pole mass to the short-distance $\overline{\rm MS}$-mass, where the latter actually enters the radiative corrections to precision observables. 
There is an on-going debate how much this theory error can be reduced in the future.
Improving the top quark mass determination still matters considering that the change from the previous value, $m_t = 173.34 \pm 0.81$~GeV,
which was only slightly higher, lowers the extracted value of $M_H$ by about 3~GeV.

If one performs a fit to the precision data only,
ignoring the direct mass measurements at the hadron colliders, one finds a somewhat higher value~\cite{PDG2016},
\begin{equation}
m_t  = 176.7 \pm 2.1~{\rm GeV} \qquad \mbox{(indirect)}.
\end{equation}
We will encounter the reason for this in Sec.~\ref{mw}. 

\begin{table}[t]
\tbl{Determinations of $M_H$}
{\begin{tabular}{@{}cc@{}}
\toprule
method & result \\
\colrule
event kinematics & $125.09 \pm 0.24$~GeV \\
Higgs branching ratios & $126.1 \pm 1.9$~GeV \\
electroweak fit & $90^{+18}_{-16}$~GeV \\
\botrule
\end{tabular}}
\label{mh}
\end{table}

\subsection{Higgs boson mass}
\label{secmh}
$M_H$ can be measured~\cite{PianoriLP2017} precisely and cleanly at the LHC by the reconstruction of Higgs decays into two photons 
and {\em via\/} intermediate states involving a real $Z$ boson and a virtual $Z^*$ into four charged leptons.
But there are two other strategies to determine $M_H$.
One invokes the Higgs decay branching ratios which for a Higgs boson of ${\cal O}(125~{\rm GeV})$ strongly vary with $M_H$.
Using ratios of the decay rates~\cite{Aad:2015zhl} into two gauge bosons, $\gamma\gamma$, $WW^*$ and $ZZ^*$,
in which some of the Higgs production uncertainties cancel, 
we find the result~\cite{PDG2016} on the second line in Table~\ref{mh}, 
in perfect agreement with the kinematically reconstructed Higgs boson mass and with an error of less than 2~GeV.
And then there is the electroweak fit itself which returns a much lower central value of $M_H$,
about $1.9~\sigma$ below those of the other two methods.

\begin{figure}[t]
\begin{center}
\includegraphics[width=360pt]{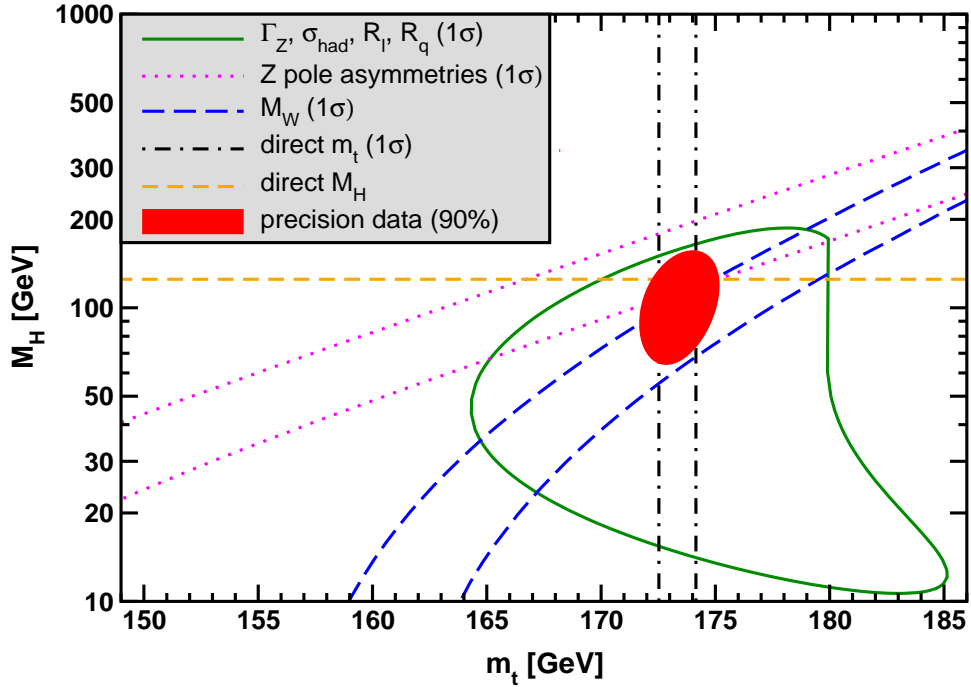}
\end{center}
\caption{One standard deviation constraints~\cite{PDG2016} on $M_H$ as functions of $m_t$ from various sets of precision data 
and the combined 90\%~CL region.}
\label{mhmt}
\end{figure}

Figure~\ref{mhmt} shows a breakdown of the constraints of the global fit.
Indicated are the direct measurements of $m_t$ and $M_H$,
the $Z$ pole asymmetries which are basically measurements of the weak mixing angle, the $W$ boson mass, $M_W$,
and a set of further $Z$ pole observables which by themselves map out a bounded part of the parameter space.
Overall, very good agreement is observed, except that $M_W$ is seen to favor lower values of $M_H$.

\section{Key Observables}
\subsection{Weak mixing angle}
\label{sec:s2w}
This section provides some details on the most important observables entering the global electroweak analysis.
The first one to discuss is the weak mixing angle, which besides $M_W$ is one of the most precise
{\em derived\/} quantities in the electroweak sector.
At the SM tree level it is given by
\begin{equation}
\sin^2\theta_W = \frac{g\prime^2}{g^2 + g\prime^2} = 1 - \frac{M_W^2}{M_Z^2}\ .
\label{s2w}
\end{equation}
Experimental results are often reported as measurements of an effective weak mixing angle, 
defined in terms of the vector and axial-vector $Z$ boson couplings to leptons, $v_\ell$ and $a_\ell$, as
\begin{equation}
\sin^2\theta_{\rm eff}^\ell \equiv \frac{1}{4} \left( 1 - \frac{v_\ell}{a_\ell} \right).
\label{s2weff}
\end{equation}

\begin{figure}[t]
\begin{center}
\includegraphics[width=360pt]{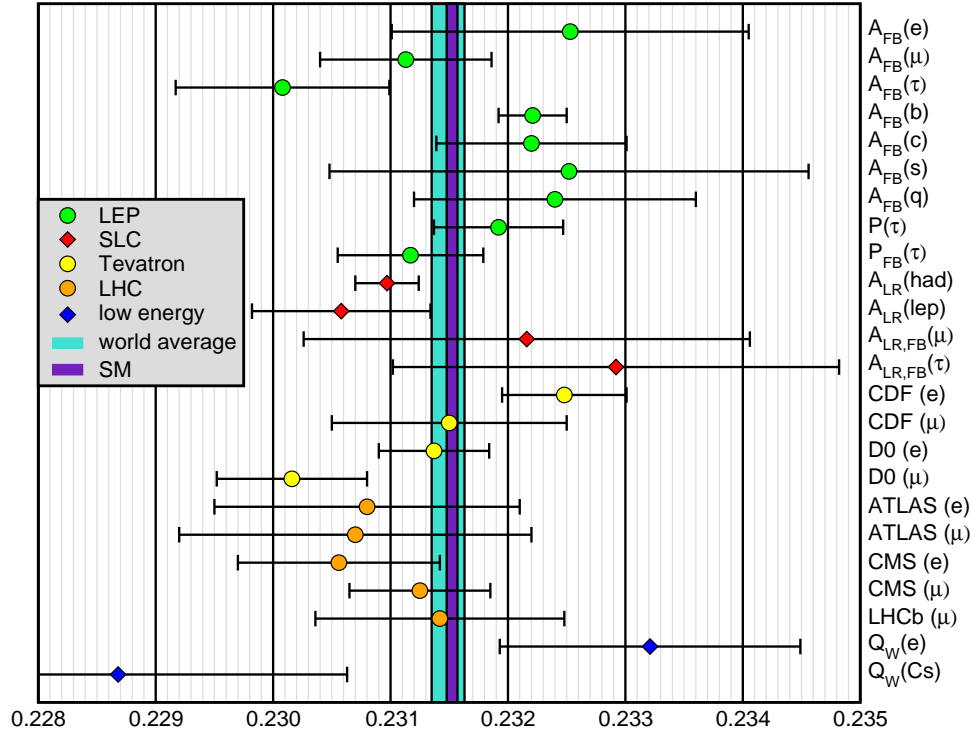}
\end{center}
\caption{Sub-\% measurements of the effective leptonic weak mixing angle.
The wider one of the two vertical bands is the world average of all measurements,
while the narrower one represents the SM prediction based on the input values for $m_t$, $M_H$, {\em etc.}}
\label{s2wsummary}
\end{figure}

Figure~\ref{s2wsummary} shows all measurements of $\sin^2\theta_{\rm eff}^\ell$ that achieved a precision of better than 1\%.
Strictly parity-violating observables are marked by diamonds, others by circles.
The first group is from the LEP Collaborations~\cite{ALEPH:2005ab}, 
featuring the forward-backward asymmetries into bottom and charm quark pairs
which are measured on the low side of the SM predictions, hence favoring values of $\sin^2\theta_W$ on the high side.
The next group is from the SLD Collaboration~\cite{ALEPH:2005ab} at the SLC, 
where the left-right polarization asymmetries into hadronic and leptonic final states both favor lower values of $\sin^2\theta_W$.

The forward-backward asymmetries for $e^+ e^-$ and $\mu^+ \mu^-$ final states have been measured 
by CDF and D\O\ at the Tevatron~\cite{sin2thetawTevatron2017}, and by ATLAS and CMS at the LHC~\cite{LiLP2017}.
The asymmetry with muon final states was also measured by the LHCb Collaboration~\cite{LiLP2017}, 
providing valuable complementarity due to the different kinematics focused on by their detector.

Finally, there are various weak charges, $Q_W$.
The weak charge of the electron, $Q_W(e)$, has been measured by the SLAC--E--158 Collaboration in polarized M\o ller scattering  
using the SLC electron beam~\cite{Anthony:2005pm}.
Qweak was the analogous experiment measuring the weak charge of the proton~\cite{Erler:2003yk}, $Q_W(p)$, in elastic $e^- p$ scattering
using the polarized electron beam at Jefferson Lab, and has been completed recently~\cite{DeconinckLP2017}.
Atomic parity violation (APV) is sensitive to the weak charges of heavy nuclei.
The most precise result was achieved in Cs in an experiment at Boulder~\cite{Bennett:1999pd}.
The very complicated atomic theory~\cite{Ginges:2003qt} is also best understood in Cs.

\begin{figure}[t]
\begin{center}
\includegraphics[width=360pt]{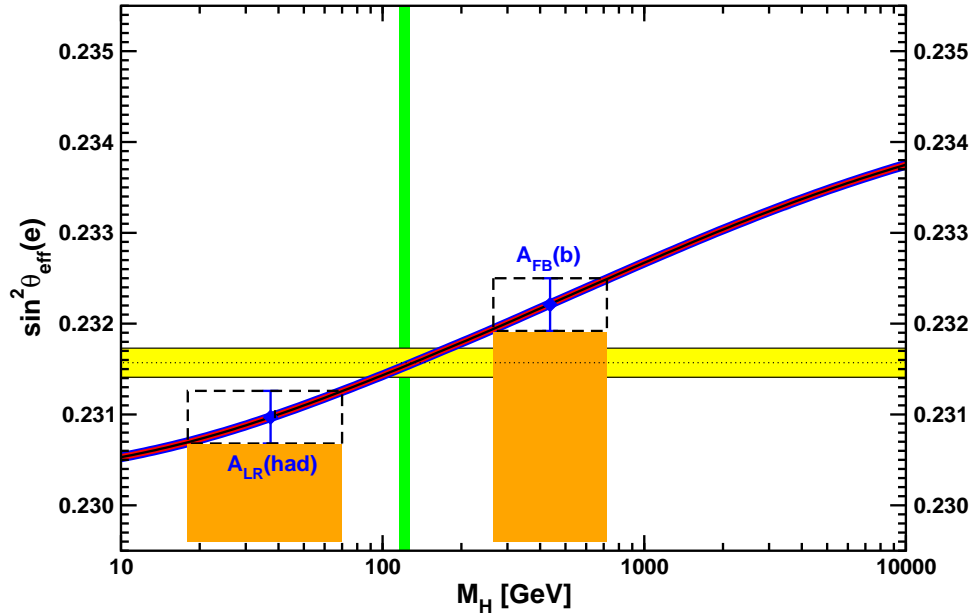}
\end{center}
\caption{Higgs mass predictions derived from $A_{LR}({\rm had})$ (SLC) and $A_{FB}(b)$ (LEP).
The former predicts Higgs boson masses of the order of tens of GeV,
while the latter prefers $M_H$ values of order hundreds of GeV.
Only the average of those and other measurements of $\sin^2\theta_W$ is truly consistent with the SM.}
\label{KKplot}
\end{figure}

Measurements of $\sin^2\theta_W$ are important for a variety of reasons even in the absence of physics beyond the SM.
They represent a unique type of test of the electroweak symmetry breaking sector,
because as summarized in Eq.~(\ref{s2w}) it is a parameter that can be written either in terms of gauge couplings,
or in terms of vector boson masses.
They also serve as a test of the still poorly studied Higgs sector,
since values of $\sin^2\theta_W$ can be translated into values of $M_H$ and then confronted with the corresponding LHC results.
And finally there is a $3~\sigma$ conflict between the most precise results from $A_{LR}({\rm had})$ and $A_{FB}(b)$
as illustrated in Fig.~\ref{KKplot}.

\begin{figure}[t]
\begin{center}
\includegraphics[width=115pt,angle=270]{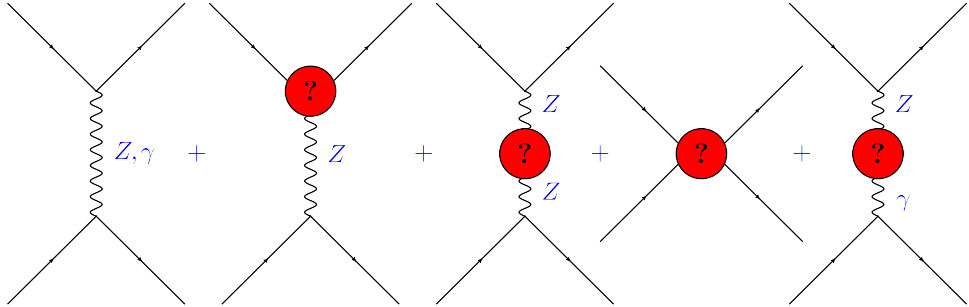}
\vspace*{0pt}
\end{center}
\caption{Sketch of how new physics may affect the extracted values of $\sin^2\theta_W$.
The first diagram represents the SM tree level, while the remaining ones represent, respectively,  
vertex corrections, oblique corrections, a non-standard four-fermion contact interaction from heavy new physics, 
and a change in the renormalization group running of $\sin^2\theta_W$ from light new physics.
One needs to vary measurement types and energy scales to disentangle these possibilities experimentally~\cite{Erler:2004cx}.}
\label{newphysics}
\end{figure}

Measurements of $\sin^2\theta_W$ are even more important in the context of new physics beyond the SM,
which can enter in very different ways as sketched in Fig.~\ref{newphysics}.

\begin{romanlist}
\item One way is through $Z$-$Z^\prime$ mixing.
If there is an extra neutral gauge boson~\cite{Langacker:2008yv,Erler:2009jh}, $Z^\prime$, 
exhibiting mass mixing with the ordinary $Z$, there may be very significant modifications of its vector couplings.
These would manifest themselves in determinations of $\sin^2\theta_W$ seemingly disagreeing with the SM.
This is also the reason why the extracted limits on $Z$-$Z^\prime$ mixing angles are very strong, and typically at the few per-mille level. 
\item Another important way is through the interpretation of the so-called oblique parameters,
which are discussed in Sec.~\ref{STU}.
\item New amplitudes may also be present, {\em e.g.\/}, from an additional $Z^\prime$ boson.
Some new four-Fermi operator could produce a measurable effect by means of interference with the photon at low momentum transfer,
but would go unnoticed in the context of measurements around the $Z$ resonance under which it would be buried. 
If one then compares on with off $Z$ pole measurements of $\sin^2\theta_W$ one may be able to isolate 
this kind of new contact interaction.
\item Finally, there is the possibility of a change in the renormalization group evolution of $\sin^2\theta_W$.
If there was a new light particle with a mass somewhere between zero and $M_Z$,
this could have an effect on the $\beta$ function~\cite{Erler:2004in} of $\sin^2\theta_W$.
\end{romanlist}

\begin{figure}[t]
\begin{center}
\includegraphics[width=360pt]{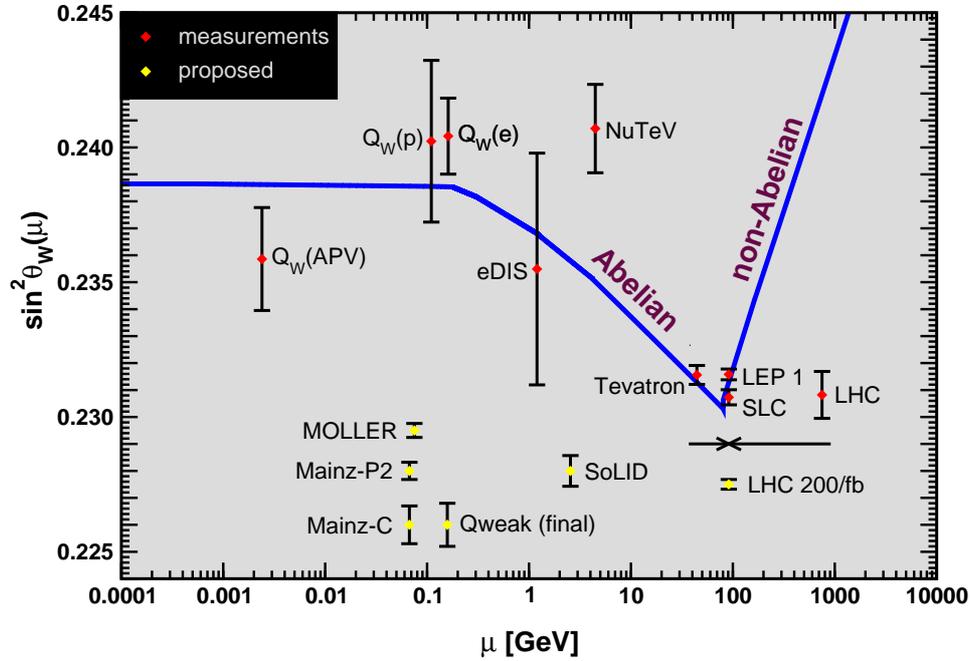}
\end{center}
\caption{Renormalization group evolution~\cite{Erler:2004in} of $\sin^2\theta_W$ in the modified minimal subtraction scheme, 
$\overline{\rm MS}$.
At the scale of $\mu = M_W$ the $\beta$-function changes sign, 
signaling the change from an effectively Abelian theory to a non-Abelian one.
Indicated are also various existing and upcoming measurements.
For more details on some of the lower energy measurements, see Ref.~\citenum{DeconinckLP2017}.
The data points around the $Z$ pole (for lack of space, the Tevatron and LHC points have been shifted horizontally)
are the averages of the individual determinations displayed in Fig.~\ref{s2wsummary},
taking into account correlated systematic errors.}
\label{runnings2w}
\end{figure}

The renormalization group running of the weak mixing angle within the SM is illustrated in Fig.~\ref{runnings2w}.
The calculation faces similar issues and problems as the calculation of the electromagnetic coupling at the $Z$ scale
in terms of $\alpha$ in the Thomson limit.
In the case of $\sin^2\theta_W$ one starts at the $Z$ pole from where the most precise measurements derive, and moves 
to lower scales to compare with the extractions from Qweak~\cite{DeconinckLP2017} or other processes involving parity-violation. 

One employs perturbative QCD wherever possible, {\em i.e.\/}, down to $\mu \approx 2$~GeV,
for which one needs precise input values of the charm and bottom quark masses.
In the region where one cannot rely on perturbation theory one can try to relate the hadronic contribution that is not calculable
from first principles to the corresponding result of $\alpha$.
While there is a part which contributes in the same way to both, $\sin^2\theta_W$ and $\alpha$, there is a complication 
because the ratio of $Z$ vector couplings to up-type and down-type quarks differs from the ratio of their electric charges,
and a flavor separation is in order.
This can be achieved to sufficient precision by constructing upper and lower bounds on the strange quark contribution~\cite{Erler:2004in}.
Another separation is needed for the singlet piece,  {\em i.e.\/}, the OZI rule violating part where one has a quark current 
connecting to a set of gluons and then connecting further to a another quark-anti-quark pair.
This piece is small but in principle introduces some additional uncertainty.
Fortunately, there is a lattice gauge theory calculation of the singlet contribution to the anomalous magnetic moment 
of the muon~\cite{KanekoLP2017} that can be adapted to this case~\cite{EF2017}.

\subsection{$W$ boson mass}
\label{mw}
\begin{figure}[t]
\begin{center}
\includegraphics[width=360pt]{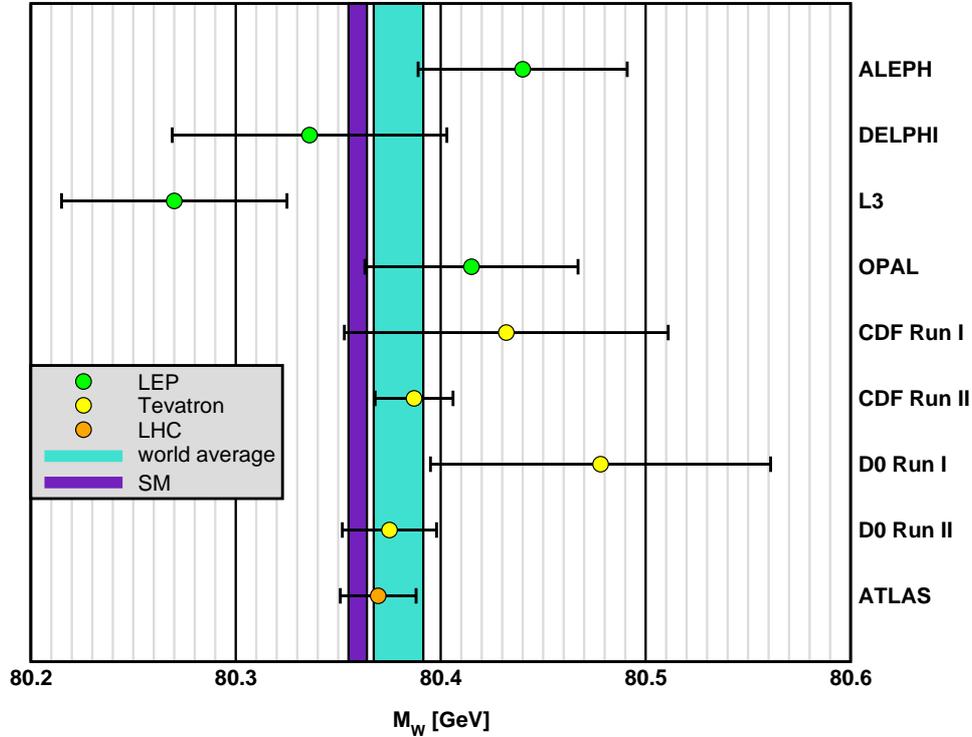}
\end{center}
\caption{Measurements of the $W$ boson mass from LEP~\cite{Schael:2013ita}, 
the Tevatron~\cite{Abazov:2012bv,Group:2012gb} and ATLAS~\cite{LiLP2017}.}
%The wider one of the two vertical bands is the world average of all measurements,
%while the narrower one represents the SM prediction based on the input values for $m_t$, $M_H$, {\em etc.}}
\label{mwsummary}
\end{figure}

\begin{figure}[t]
\begin{center}
\includegraphics[width=360pt]{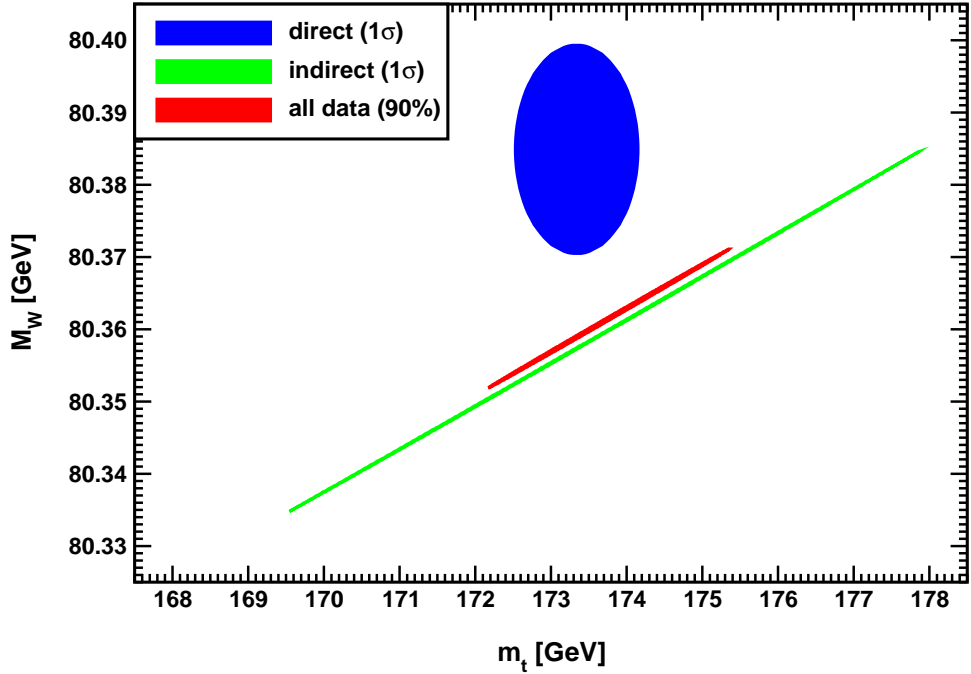}
\end{center}
\caption{Constraints~\cite{PDG2016} on $M_W$ as functions of $m_t$.
The blue ellipse represents the direct measurements, while the long green line is the $1~\sigma$ contour from all other data.
The shorter red line is the combination of all data at the 90\%~CL.
The approximately $2~\sigma$ SM deviation in $M_W$ is apparent.}
\label{mwmt}
\end{figure}

Another key observable is $M_W$ where the status is almost the opposite from $\sin^2\theta_W$.
As can be seen in Fig.~\ref{mwsummary}, the most precise measurements are in perfect agreement with each other,
but the central values of all available measurements except for DELPHI and L3 are higher than the SM prediction.
As a result, the world average is off by about two standard deviations.
This is also transparent from Fig.~\ref{mwmt}.

There is a very interesting interpretation of an enhanced $M_W$ within the Minimal Supersymmetric Standard Model (MSSM).
While the size of a possible shift in $M_W$ is not clearly predicted,  
the overall sign of the MSSM contributions~\cite{Heinemeyer:2013dia} is expected to increase $M_W$ relative to the SM prediction,
in agreement with what is currently seen.
This is regardless of whether the boson that the LHC has discovered was the lighter or the heavier of the two CP-even Higgs eigenstates
that are present in the MSSM, but the latter case is much more constrained~\cite{Bechtle:2016kui}.

\subsection{Anomalous magnetic moment of the muon}
The final key observable is the anomalous magnetic moment of the muon,
\begin{equation}
a_\mu \equiv \frac{g_\mu - 2}{2} = \frac{\alpha}{2\pi} + {\cal O}(\alpha^2)\ .
\label{amu}
\end{equation}
It has been measured by the Muon $g-2$ (BNL--E--821) Collaboration~\cite{Bennett:2006fi},
\begin{equation}
\label{amuexp}
a_\mu =  (1165920.91 \pm 0.63) \times 10^{-9} ,
\end{equation}
and there will be a follow-up experiment at Fermilab~\cite{Grange:2015fou}, 
as well as a conceptually different experiment at J-PARC~\cite{JPARC2011}, 
each attempting to reduce the experimental uncertainty by a factor of about four.
The SM prediction,
\begin{equation}
a_\mu =  (1165917.63 \pm 0.46) \times 10^{-9} ,
\end{equation}
deviates by $4.2~\sigma$ from Eq.~(\ref{amuexp}), but there are some issues regarding the SM value.

\begin{figure}[t]
\begin{center}
\parbox{2.1in}{\includegraphics[width=150pt,height=146pt]{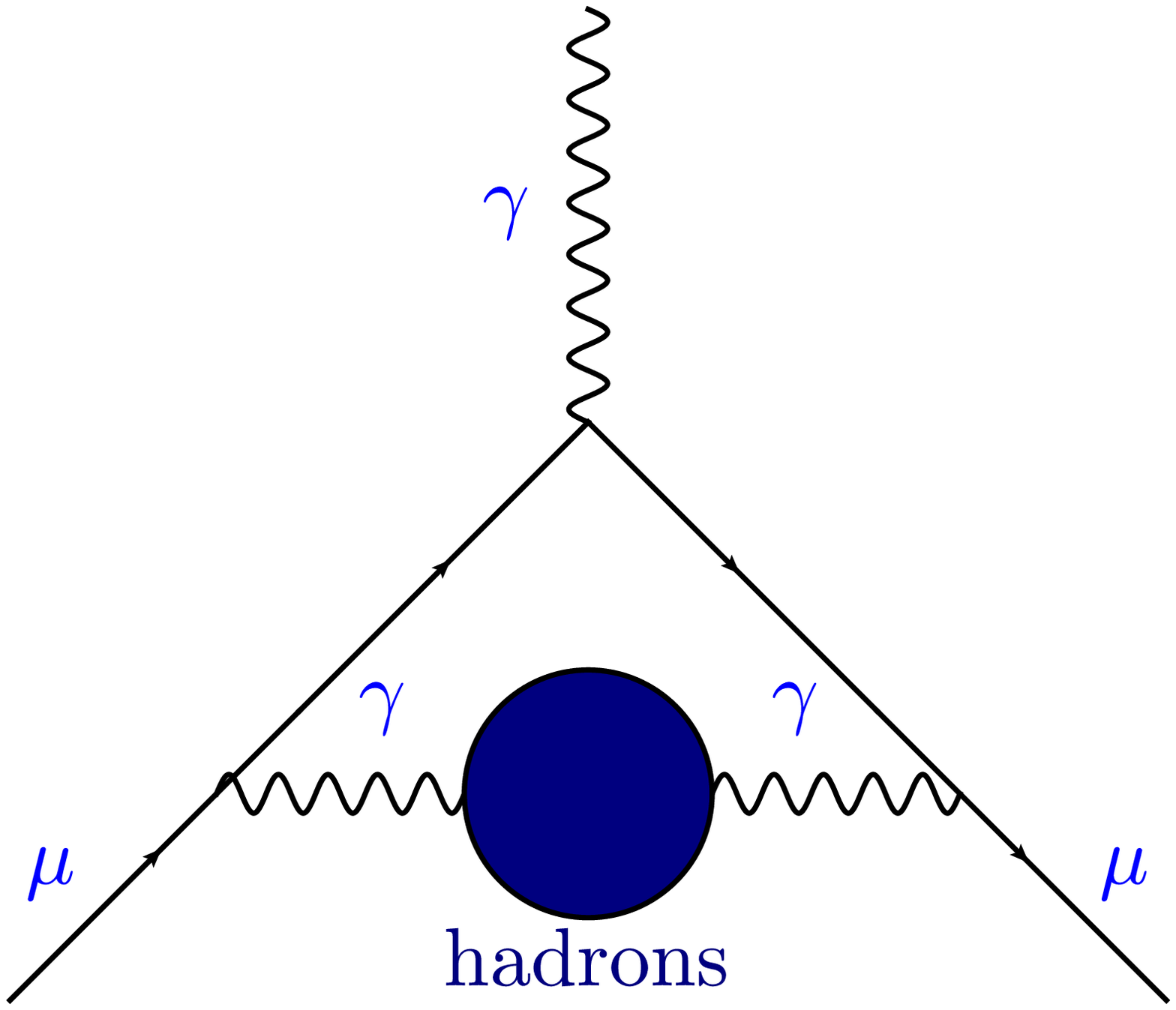}\vspace{8pt}\figsubcap{a}}
\hspace*{25pt}\vspace{5pt}
\parbox{2.1in}{\includegraphics[width=150pt]{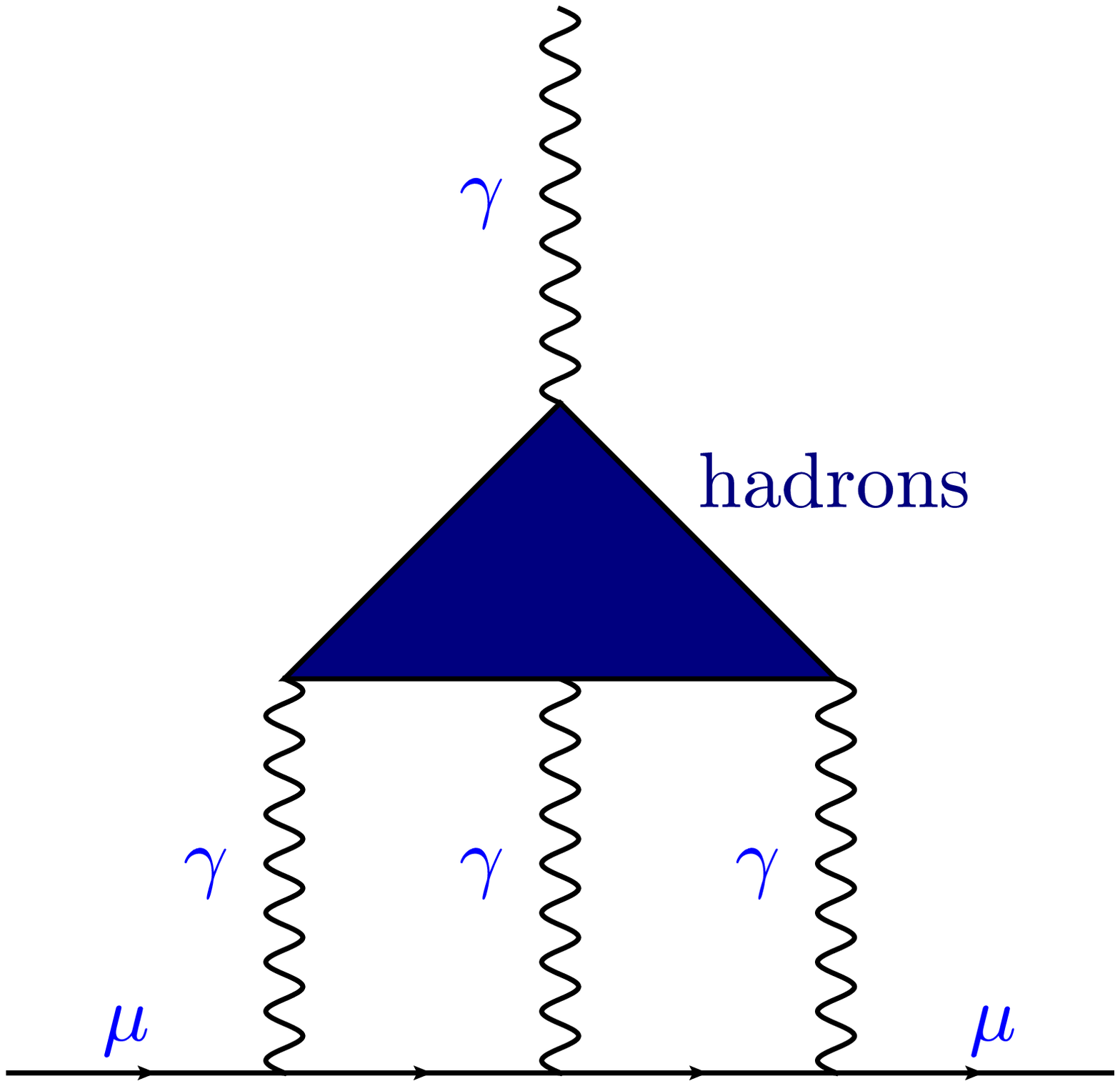}\figsubcap{b}}
\caption{Hadronic contributions to the anomalous magnetic moment of the muon.
(a) Two-loop vacuum polarization contribution.
(b) Three-loop light-by-light scattering contribution.} 
\label{hadroniceffects}
\end{center}
\end{figure}

One is the hadronic vacuum polarization contribution which enters first at the two-loop level and is
depicted in Fig.~\ref{hadroniceffects}(a).
Again, one can use perturbative QCD for part of the effect~\cite{Erler:2000nx},
and just as in Sec.~\ref{sec:s2w} one needs values of the bottom and charm quark masses as inputs.
Here most of the contribution is from the lower energy hadronic regime,
where one has to resort to data from experiments or lattice gauge theory simulations.
Currently, experimental data come from three different types of sources:
\begin{romanlist}[(iii)]
\item energy scans measuring $e^+ e^-$ annihilation cross sections into hadronic final states;
\item radiative returns~\cite{Rodrigo:2001kf} from resonances such as the $\phi$ or the $\Upsilon(4S)$ 
(also produced in $e^+ e^-$ annihilation),
{\em i.e.\/}, decays where the energy is shared between the hadronic system and an additional photon 
(this method is dominated by systematic and theoretical uncertainties);
\item spectral functions of $\tau^\pm \rightarrow \nu_\tau \pi^\pm \pi^0$ decays,
which are related by an isospin rotation to $e^+ e^- \rightarrow \pi^+ \pi^-$. 
Final states with four pions can also be used.
The method assumes isospin symmetry and isospin violating effects have to be corrected for introducing an extra uncertainty.
\end{romanlist}
The experimental results from $e^+ e^-$ annihilation and $\tau$ decays are discussed in detail in Ref.~\citenum{Davier:2009ag}.
In each of the two classes, the results are in very good agreement with each other, but the analysis based on $\tau$ decays
showed a smaller deviation from the SM which prefers higher values of the hadronic contribution to $a_\mu$.
However, a few years ago it was pointed out~\cite{Jegerlehner:2011ti} 
that another isospin-breaking effect originating from $\gamma$-$\rho$ mixing needs to be included.
Ref.~\citenum{Jegerlehner:2011ti} obtained a parameter-free prediction for the size this correction
and applying it moves the $\tau$ based analysis into very good agreement with the $e^+ e^-$ data.
Averaging them together implies a larger deviation compared to considering the $e^+ e^-$ data alone.

There are interesting proposals to reduce the hadronic vacuum polarization uncertainty in $a_\mu$ in the future.
They involve to measure the vacuum polarization in the space-like region from Bhabha scattering~\cite{Calame:2015fva} 
or from $\mu e$ scattering~\cite{Abbiendi:2016xup}.
Such an experiment would measure the running of $\alpha$, 
and the corresponding contribution to $a_\mu$ would be achieved by means of the convolution~\cite{Lautrup:1971jf},
\begin{equation}
a_\mu = \frac{\alpha}{\pi} \int_0^1 dx (1-x) \Delta\alpha \left( \frac{x^2 m_\mu^2}{x-1} \right).
\end{equation}

Another difficult issue is the hadronic light-by-light contribution shown in Fig.~\ref{hadroniceffects}(b).
Entering first at the three-loop level, it is much smaller than that of the hadronic vacuum polarization.
However, its uncertainty is nevertheless numerically comparable and crudely estimated to be roughly 30\% of itself. 
The reason is that there are not have enough data to constrain the associated dispersion integral
which in this case is two-dimensional.  
Fortunately, there are promising efforts from lattice gauge theory~\cite{KanekoLP2017}. 
A recent calculation of the quark-connected and leading disconnected contributions~\cite{Blum:2016lnc} 
achieved a 25\% statistical precision while the systematics is currently under investigation. 
There is a valuable and encouraging cross-check in which the muonic light-by-light contribution 
has been simulated on the lattice, and the result agrees within about 2\% with perturbative QED~\cite{Blum:2017cer}.

There are also recent lattice results~\cite{DellaMorte:2017dyu} on the vacuum polarization contribution.
Here the uncertainty could be reduced to 6\%, and sub-\% precision may be achievable in the not too distant future.
Note that about half a percent accuracy would be needed to be competitive with the dispersion result.

\section{New Physics Implications}
\subsection{Oblique physics beyond the Standard Model}
\label{STU}
The oblique parameters describe radiative corrections to the vector boson two-point correlation functions,
and in one specific formalism~\cite{Peskin:1991sw} are called the $S$, $T$ and $U$ parameters.
They may be defined to describe new physics only, such that $S = T = U = 0$ in the SM.
To extract them from the data one needs measurements of $\sin^2\theta_W$, $M_W$, 
as well as at least one other derived observable, such as the $Z$~boson width, $\Gamma_Z$, or a low-energy neutral current observable.

The oblique parameters are easily affected by most models of physics beyond the SM,
especially those addressing naturalness issues of the Higgs potential (the hierarchy problem).
The oblique approximation neglects possible additional new physics effects, such as direct contributions to fermion couplings.
Even where this is not a reasonable approximation, it still provides a valuable reference case.

The $T$ parameter breaks the accidental (custodial) $SO(4)$ symmetry present in the Higgs potential.
Its effects are indistinguishable from tree-level new physics corrections to the $\rho_0$ parameter,
defined as the ratio of the neutral-current to charged-current interaction strengths.

A new multiplet of heavy {\em degenerate\/} chiral fermions would produce a constant contribution,
\begin{equation}
\Delta S = \sum_i \frac{N_C^i}{3\pi} (t_{3L}^i - t_{3R}^i)^2,
\end{equation}
to the $S$ parameter, where $N_C^i$ is the color factor of multiplet $i$.
{\em E.g.\/}, an additional complete and degenerate fermion generation or mirror generation would contribute 
\[ \Delta S = \frac{2}{3\pi} \approx 0.21\ .\]

Another way to think about $S$ and $T$ is that they correspond to dimension six operators in the SM effective field theory,
in which one supplements the SM with additional gauge-invariant, but non-renormalizable operators.
The $U$ parameter corresponds to a combination of dimension eight operators,
so that $U$ is expected to be more suppressed than $S$ and $T$.
This expectation is indeed borne out in concrete models.

\subsection{Non-degenerate doublets}
A simple example is given by a set of extra non-degenerate doublets 
which would contribute to $\rho_0 = 1 + \alpha T$ as~\cite{Veltman:1977kh},
\begin{equation}
\Delta \rho_0 = \frac{G_F}{8\sqrt{2} \pi^2} \sum_i N_C^i \Delta m_i^2\ ,
\end{equation}
where $\Delta m_i^2$ is not simply the difference of the masses squared of the two members of each doublet,
but rather a more complicated function with the property that $\Delta m_i^2 \geq (m_1 - m_2)^2$.
Thus, it is a positive definite function, and despite appearances, {\em there is\/} decoupling in this formula.
At first sight one could have the impression that a very heavy doublet with, say, Planck scale masses and a mass difference of
electroweak size, could give a measurable effect even at present day colliders.
However, this is not the case, because in actual models 
in turns out that $\Delta m_i^2$ itself will experience a see-saw type of suppression and will tend to zero,
consistent with the interpretation of $T$ as a dimension six operator.

With the latest results on $\sin^2\theta_W$, {\em etc.\/}, I now find,
\begin{equation}
\rho_0 = 1.00039 \pm 0.00019\ ,
\end{equation}
which differs from the SM prediction, $\rho_0 = 1$, by 2.0~standard deviations.
It is amusing that this implies a {\em non-vanishing new physics contribution\/} at the 90\%~CL,
for which I find the range,
\begin{equation}
(15\mbox{ GeV})^2 \leq \sum_i \frac{N_C^i}{3} \Delta m_i^2 \leq (47\mbox{ GeV})^2.
\label{deltam2}
\end{equation}
Note, that the future Circular Electron Positron Collider (CEPC) under consideration in China, 
could measure $\rho_0$ with a precision of $8 \times 10^{-5}$,
and assuming that the central value would not change from today,
the quantity in Eq.~(\ref{deltam2}) would be discovered to be non-zero at the $5~\sigma$ level.

\begin{figure}[t]
\begin{center}
\includegraphics[width=360pt]{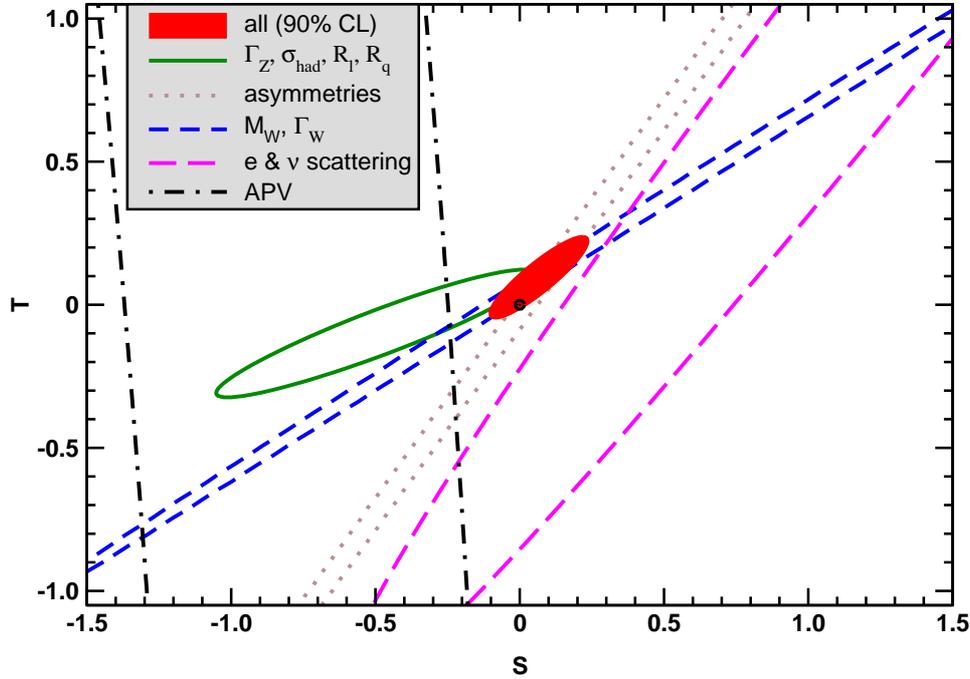}
\end{center}
\caption{Constraints~\cite{PDG2016} on $T$ as functions of $S$.}
\label{ST}
\end{figure}

\subsection{The S and T parameters}
The result of a simultaneous fit to $S$ and $T$ is shown in Fig~\ref{ST}.
Notice that with the exception of the constraint from atomic parity violation (which is from Cs and Tl experiments)
all constraints have a similar positive slope as they are all related to $\sin^2\theta_W$.  
The various classes of constraints and the combined fit are in reasonable agreement with the SM prediction, $S = T = 0$,
but at the best fit values,
\begin{align}
S = 0.06 \pm 0.08\ , \\
T = 0.09 \pm 0.06\ ,
\label{STfit}
\end{align}
the fit is moderately better, showing a decrease of $\Delta\chi^2 = -4.0$ compared to the SM global fit.
It would be interesting to improve the precision in such a fit.
This could be done, for example, at the CEPC which could achieve uncertainties of $\pm 0.014$ and $\pm 0.017$ in $S$ and $T$, respectively.

The $S$ parameter by itself rules out QCD-like technicolor models, and only much more complicated models may still be viable.
$S$ also incontrovertibly rules out a degenerate fourth fermion generation,
and even a non-degenerate one is highly disfavored~\cite{Erler:2010sk}.

\begin{figure}[t]
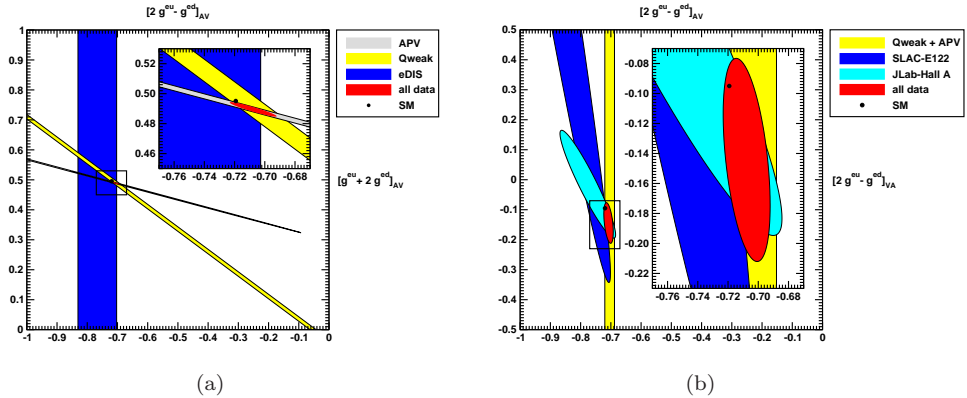

\begin{center}
\hspace*{-25pt}
\parbox{2.1in}{\includegraphics[width=162pt]{contour1.eps}\vspace{8pt}\figsubcap{a}}
\hspace*{25pt}\vspace{5pt}
\parbox{2.1in}{\includegraphics[width=175pt]{contour2.eps}\vspace{8pt}\figsubcap{b}}
\caption{Constraints on parity-violating four-fermion operators~\cite{Erler:2014fqa}.
Notice that the abscissa is identical in the two figures.
(a) Operators containing the axial-vector electron bilinear $\overline e \gamma^5\gamma^\mu e$.
(b) Operators containing the charge-weighted quark combinations.} 
\label{contours}
\end{center}
\end{figure}

\subsection{Non-oblique parameters}
As already mentioned in Sec.~\ref{sec:s2w}, there is a long-standing SM deviation in the forward-backward asymmetry, $A_{FB}(b)$,
as measured at LEP.
Interpreting $A_{FB}(b)$ instead of a measurement of $\sin^2\theta_W$ as a measurement of the flavor-dependent 
form factor $\Delta\kappa_b$ multiplying the weak mixing angle relevant for $b$ quarks, and performing a fit simultaneously with another 
form factor $\Delta\rho_b$ which can be thought of as the $\rho_0$ parameter for $b$ quarks, we find~\cite{PDG2016},
\begin{align}
\Delta\rho_b = 0.056 \pm 0.020\ , \\
\Delta\kappa_b = 0.182 \pm 0.068\ .
\label{FFbquark}
\end{align}
This represents a $2.7~\sigma$ deviation from the SM prediction $\Delta\rho_b = \Delta\kappa_b = 0$, 
and is driven by the deviation in $A_{FB}(b)$.
However, it is difficult to explain this deviation in terms of new physics
without also shifting the $Z$ boson branching ratio into $b$~quarks, $R_b$, which is in reasonable agreement with the SM.

\begin{figure}[t]
\begin{center}
\includegraphics[width=360pt]{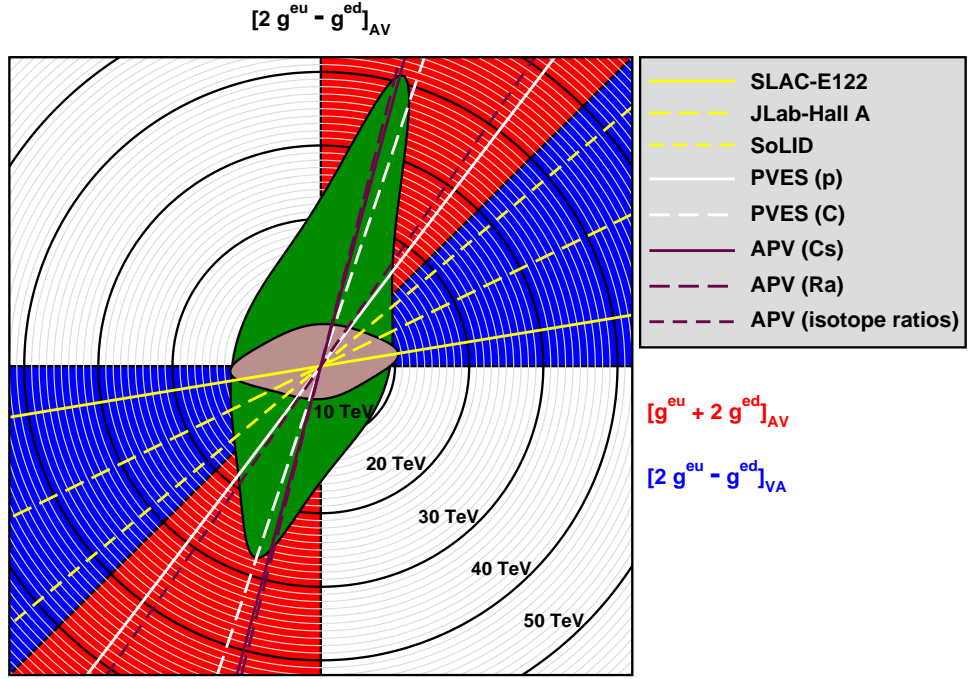}
\end{center}
\caption{Reach to new physics scales~\cite{Erler:2014fqa} from the four-fermion operator constraints in Fig.~\ref{contours}.
The blue segment is accessible to DIS experiments (yellow lines) and defines a plane 
(containing the brown 95\% CL exclusion contour) perpendicular to the plane containing the red segment, 
the green contour and the white and maroon lines. 
Thus, the two planes are subspaces of a three-dimensional parameter space which intersect along the horizontal direction. 
The lines indicate the coupling combinations of the various experiments relative to the common horizontal direction}
\label{scales}
\end{figure}

The CEPC may be able to achieve precisions of $\pm 0.005$ and $\pm 0.007$ in $\Delta\rho_b$ and $\Delta\kappa_b$, respectively.
Note that the results in Eq.~(\ref{FFbquark}) are essentially independent of $S$, $T$, and $U$, meaning
that whether you allow them to also vary in the fits or fix them to the SM, makes little difference in these extractions.

\subsection{Compositeness scales from low energies}
From polarized electron-proton~\cite{DeconinckLP2017}, M\o ller, and deep inelastic scattering (DIS), as well as APV,
one can obtain constraints in a three-dimensional coupling space of parity-violating four-fermion operators,
appearing in the effective Lagrangian~\cite{Erler:2013xha},
\begin{align}
\label{LeqNC}
{\cal L}_{\rm NC}^{\, e q} = &- {2 \over v^2} {\overline e \gamma^5\gamma^\mu e \over 2}
\left[ g_{AV}^{\, e u} {\overline{u} \gamma_\mu u \over 2} + 
g_{AV}^{\, e d} {\overline d \gamma_\mu d \over 2} \right] \nonumber \\
&- {2 \over v^2} {\overline e \gamma^\mu e \over 2} 
\left[ g_{VA}^{\, e u} {\overline{u} \gamma^5 \gamma_\mu u \over 2} + 
g_{VA}^{\, e d} {\overline d \gamma^5 \gamma_\mu d \over 2} \right],
\end{align}
where $v = (\sqrt{2} G_F)^{-1/2} = 246.22$~GeV is the Higgs vacuum expectation value. 
The SM tree-level relations for the real-valued coefficients $g_{AV}^{\, e q}$ and $g_{VA}^{\, e q}$ are given by
\begin{align}
\label{gAVeud}
g_{AV}^{\, e u} \equiv& \cos^2\theta_W g_A^e g_V^u = - {1\over 2} + {4\over 3} \sin^2\theta_W, \\
\qquad\qquad
g_{AV}^{\, e d} \equiv& \cos^2\theta_W g_A^e g_V^d = \phantom- {1\over 2} - {2\over 3}  \sin^2\theta_W, \\
\label{gVAeud}
g_{VA}^{\, e u}  \equiv& \cos^2\theta_W g_V^e g_A^u = - {1\over 2} + 2 \sin^2\theta_W, \\
\qquad\qquad
g_{VA}^{\, e d}  \equiv& \cos^2\theta_W g_V^e g_A^d =  \phantom- {1\over 2} - 2 \sin^2\theta_W.
\end{align}
The results and their combinations are shown in Fig.~\ref{contours}.
For Fig.~\ref{scales} these have been translated into sensitivities to new physics scale.
In the future, such scales may reach up to about 50~TeV provided one assumes the strong coupling case, 
as {\em e.g.\/}, in compositeness models.

\section{Conclusions}
To summarize, the SM is almost 50~years old, but in remarkable health.
It is over-constrained, where derived quantities like $\sin^2\theta_W$, $M_W$, $g_\mu - 2$, and weak charges 
have been both computed and measured.
The precision in future measurements of $\sin^2\theta_W$, $M_W$, $Q_W(e)$ and $Q_W(p)$ will challenge theory,
and a major global effort is needed to keep the theory uncertainties well below the projected experimental ones.
Contact interactions derived from comparing future measurements of $\sin^2\theta_W$ at low energies~\cite{Kumar:2013yoa}
with determinations near the $Z$ pole will be able to test new physics scales up to around 50~TeV.

Currently, the indirectly determined Higgs boson mass from the global fit is $1.9~\sigma$ below the directly reconstructed mass,
while the $\rho_0$ parameter is $2.0~\sigma$ high.
Both observations can be traced to the measured $M_W$ which is above the SM prediction.

\section*{Acknowledgments}
I would like to thank the organizers for the kind invitation to a very enjoyable symposium.
This work is supported by CONACyT (M\'exico) project 252167--F.

\bibliographystyle{ws-procs961x669}

\begin{thebibliography}{99}

\bibitem{Weinberg:1967tq} 
S.~Weinberg, 
A Model of Leptons,
{\em Phys.\ Rev.\ Lett.}\  {\bf 19}, 1264 (1967).
%%CITATION = doi:10.1103/PhysRevLett.19.1264;%%

\bibitem{Erler:2012uu} 
J.~Erler,
Weighing in on the Higgs, 
arXiv:1201.0695 [hep-ph].
%%CITATION = ARXIV:1201.0695;%%

\bibitem{Patrignani:2016xqp} 
Particle Data Group: C.~Patrignani {\it et al.}, 
Review of Particle Physics, \\
{\em Chin.\ Phys.\ C} {\bf 40}, 100001 (2016).
%%CITATION = doi:10.1088/1674-1137/40/10/100001;%%
  
\bibitem{Webber:2010zf} 
MuLan Collaboration: D.~M.~Webber {\it et al.},
Measurement of the Positive Muon Lifetime and Determination of the Fermi Constant to Part-per-Million Precision, \\
{\em Phys.\ Rev.\ Lett.}\ {\bf 106}, 041803 (2011),
arXiv:1010.0991 [hep-ex].
%%CITATION = doi:10.1103/PhysRevLett.106.041803, 10.1103/PhysRevLett.106.079901;%%
  
\bibitem{ALEPH:2005ab} 
ALEPH, DELPHI, L3, OPAL, and SLD Collaborations, LEP Electroweak Working Group, and SLD Electroweak and Heavy Flavour Groups: 
S.~Schael {\it et al.},  \\
Precision electroweak measurements on the $Z$ resonance, \\
{\em Phys.\ Rept.}\ {\bf 427}, 257 (2006),
hep-ex/0509008.
%%CITATION = doi:10.1016/j.physrep.2005.12.006;%%  
  
\bibitem{PianoriLP2017}
Elisabetta Pianori, 
Higgs boson measurements in diboson decay modes, \\
these proceedings.
  
\bibitem{PDG2016}  
J.~Erler and A.~Freitas, 
Electroweak Model and Constraints on New Physics, \\
pp.~151--171 in Ref.~\citenum{Patrignani:2016xqp}.
 
\bibitem{LeDiberder:1992jjr} 
F.~Le Diberder and A.~Pich,
The perturbative QCD prediction to $R_\tau$ revisited, \\
{\em Phys.\ Lett.\ B} {\bf 286}, 147 (1992).
%%CITATION = doi:10.1016/0370-2693(92)90172-Z;%%
  
\bibitem{Boito:2012cr} 
D.~Boito, M.~Golterman, M.~Jamin, A.~Mahdavi, K.~Maltman, J.~Osborne and S.~Peris,\\
An Updated determination of $\alpha_s$ from $\tau$ decays, \\
{\em Phys.\ Rev.\ D} {\bf 85}, 093015 (2012),
arXiv:1203.3146 [hep-ph].
%%CITATION = doi:10.1103/PhysRevD.85.093015;%%
  
\bibitem{Boito:2014sta} 
D.~Boito, M.~Golterman, K.~Maltman, J.~Osborne and S.~Peris, \\
Strong coupling from the revised ALEPH data for hadronic $\tau$ decays,\\
{\em Phys.\ Rev.\ D} {\bf 91}, 034003 (2015),
arXiv:1410.3528 [hep-ph].
%%CITATION = doi:10.1103/PhysRevD.91.034003;%%  
  
\bibitem{YamazakiLP2017}
Yuji Yamazaki, 
Top-Quark Measurements,
these proceedings.
  
\bibitem{Nisius:2017ppa} 
R.~Nisius,
Measurements of the top quark mass with the ATLAS detector,
arXiv:1709.09845 [hep-ex].

\bibitem{Spannagel:2016cqt} 
S.~Spannagel,
Top quark mass measurements with the CMS experiment at the LHC,
{\em PoS DIS} {\bf 2016}, 150 (2016),
arXiv:1607.04972 [hep-ex].
%%CITATION = ARXIV:1607.04972;%%
  
\bibitem{TevatronElectroweakWorkingGroup:2016lid} 
CDF and D\O\ Collaborations, and Tevatron Electroweak Working Group: \\ T. Aaltonen {\it et al.},
Combination of CDF and D0 results on the mass of the top quark using up $9.7\:{\rm fb}^{-1}$ at the Tevatron,
arXiv:1608.01881 [hep-ex].
%%CITATION = ARXIV:1608.01881;%%

\bibitem{Erler:2015nsa} 
J.~Erler,
On the Combination Procedure of Correlated Errors, \\
{\em Eur.\ Phys.\ J.\ C} {\bf 75}, 453 (2015),
arXiv:1507.08210 [physics.data-an].
%%CITATION = doi:10.1140/epjc/s10052-015-3688-y;%%

\bibitem{Marquard:2016dcn} 
P.~Marquard, A.~V.~Smirnov, V.~A.~Smirnov, M.~Steinhauser and D.~Wellmann, \\
$\overline{\rm MS}$-on-shell quark mass relation up to four loops in QCD and a general SU$(N)$ gauge group,
{\em Phys.\ Rev.\ D} {\bf 94}, 074025 (2016),
arXiv:1606.06754 [hep-ph].
%%CITATION = doi:10.1103/PhysRevD.94.074025;%%
  
\bibitem{Aad:2015zhl} 
ATLAS and CMS Collaborations: G.~Aad {\it et al.},
Combined Measurement of the Higgs Boson Mass in $pp$ Collisions at $\sqrt{s}=7$ and 8 TeV with the ATLAS and CMS Experiments,
{\em Phys.\ Rev.\ Lett.}\  {\bf 114}, 191803 (2015),
arXiv:1503.07589 [hep-ex], and \\
\url{cds.cern.ch/record/2052552/files/ATLAS-CONF-2015-044.pdf}.
%%CITATION = doi:10.1103/PhysRevLett.114.191803;%%

\bibitem{sin2thetawTevatron2017}
CDF and D\O\ Collaborations and the Tevatron Electroweak Working Group, \\
Tevatron combination of the effective leptonic electroweak mixing angles (2017), \\
\url{https://tevewwg.fnal.gov/wz/sw2eff17/drafts/Fermilab_Conf_17_201_E.pdf}.
  
\bibitem{LiLP2017}
Qiang Li,
EW Measurements from LHC and Past Experiments,
these proceedings.
  
\bibitem{Anthony:2005pm} 
SLAC--E--158 Collaboration: P.~L.~Anthony {\it et al.},
Precision measurement of the weak mixing angle in M\o ller scattering,
{\em Phys.\ Rev.\ Lett.}\ {\bf 95}, 081601 (2005),
hep-ex/0504049.
%%CITATION = doi:10.1103/PhysRevLett.95.081601;%%
  
\bibitem{Erler:2003yk} 
J.~Erler, A.~Kurylov and M.~J.~Ramsey-Musolf,
The Weak charge of the proton and new physics,
{\em Phys.\ Rev.\ D} {\bf 68}, 016006 (2003),
hep-ph/0302149.
%%CITATION = doi:10.1103/PhysRevD.68.016006;%%
    
\bibitem{DeconinckLP2017}
Wouter Deconinck,
Precision Measurements with Leptons and Kaons and Nuclei,\\
these proceedings.
  
\bibitem{Bennett:1999pd} 
S.~C.~Bennett and C.~E.~Wieman,
Measurement of the $6S \rightarrow 7S$ transition polarizability in atomic cesium and an improved test of the Standard Model, \\
{\em Phys.\ Rev.\ Lett.}\  {\bf 82}, 2484 (1999),
hep-ex/9903022.
%%CITATION = doi:10.1103/PhysRevLett.82.2484;%%
  
\bibitem{Ginges:2003qt} 
J.~S.~M.~Ginges and V.~V.~Flambaum,
Violations of fundamental symmetries in atoms and tests of unification theories of elementary particles, \\
{\em Phys.\ Rept.}\ {\bf 397}, 63 (2004),
physics/0309054.
%%CITATION = doi:10.1016/j.physrep.2004.03.005;%%
  
\bibitem{Erler:2004cx} 
J.~Erler and M.~J.~Ramsey-Musolf,
Low energy tests of the weak interaction, \\
{\em Prog.\ Part.\ Nucl.\ Phys.}\ {\bf 54}, 351 (2005),
hep-ph/0404291.
%%CITATION = doi:10.1016/j.ppnp.2004.08.001;%%
  
\bibitem{Langacker:2008yv} 
P.~Langacker,
The Physics of Heavy $Z^\prime$ Gauge Bosons, \\
{\em Rev.\ Mod.\ Phys.}\ {\bf 81}, 1199 (2009),
arXiv:0801.1345 [hep-ph].
%%CITATION = doi:10.1103/RevModPhys.81.1199;%%

\bibitem{Erler:2009jh} 
J.~Erler, P.~Langacker, S.~Munir and E.~Rojas,
Improved Constraints on Z-prime Bosons from Electroweak Precision Data, \\
{\em JHEP} {\bf 0908}, 017 (2009),
arXiv:0906.2435 [hep-ph].
%%CITATION = doi:10.1088/1126-6708/2009/08/017;%%
  
\bibitem{Erler:2004in} 
J.~Erler and M.~J.~Ramsey-Musolf,
The Weak mixing angle at low energies, \\
{\em Phys.\ Rev.\ D} {\bf 72}, 073003 (2005),
hep-ph/0409169.
%%CITATION = doi:10.1103/PhysRevD.72.073003;%%
  
\bibitem{KanekoLP2017}
Takashi Kaneko,
LQCD: Flavor Physics and Spectroscopy,
these proceedings.
  
\bibitem{EF2017}
J.~Erler and R.~Ferro-Hern\'andez, \\
Reduced Uncertainties in the Weak Mixing Angle at Low Energies,
in preparation.

\bibitem{Schael:2013ita} 
ALEPH, DELPHI, L3, and OPAL Collaborations and LEP Electroweak Working Group: S.~Schael {\it et al.},
Electroweak Measurements in Electron-Positron Collisions at W-Boson-Pair Energies at LEP, \\
{\em Phys.\ Rept.}\ {\bf 532}, 119 (2013),
arXiv:1302.3415 [hep-ex].
%%CITATION = doi:10.1016/j.physrep.2013.07.004;%%  

\bibitem{Abazov:2012bv} 
D\O\ Collaboration: V.~M.~Abazov {\it et al.},
Measurement of the W Boson Mass with the D0 Detector,
{\em Phys.\ Rev.\ Lett.}\ {\bf 108}, 151804 (2012),
arXiv:1203.0293 [hep-ex].
%%CITATION = doi:10.1103/PhysRevLett.108.151804;%%
  
\bibitem{Group:2012gb} 
CDF and D\O\ Collaborations and the Tevatron Electroweak Working Group, \\
2012 Update of the Combination of CDF and D0 Results for the Mass of the W Boson,
arXiv:1204.0042 [hep-ex].
%%CITATION = ARXIV:1204.0042;%%

\bibitem{Heinemeyer:2013dia} 
S.~Heinemeyer, W.~Hollik, G.~Weiglein and L.~Zeune, \\
Implications of LHC search results on the W boson mass prediction in the MSSM,
{\em JHEP} {\bf 1312}, 084 (2013),
arXiv:1311.1663 [hep-ph].
%%CITATION = doi:10.1007/JHEP12(2013)084;%%

\bibitem{Bechtle:2016kui} 
P.~Bechtle, H.~E.~Haber, S.~Heinemeyer, O.~St\aa l, T.~Stefaniak, G.~Weiglein, L.~Zeune,
The Light and Heavy Higgs Interpretation of the MSSM, \\
{\em Eur.\ Phys.\ J.\ C} {\bf 77}, 67 (2017),
arXiv:1608.00638 [hep-ph].
%%CITATION = doi:10.1140/epjc/s10052-016-4584-9;%%
  
\bibitem{Bennett:2006fi} 
Muon g-2 Collaboration: G.~W.~Bennett {\it et al.},
Final Report of the Muon E821 Anomalous Magnetic Moment Measurement at BNL, \\
{\em Phys.\ Rev.\ D} {\bf 73}, 072003 (2006),
hep-ex/0602035.
%%CITATION = doi:10.1103/PhysRevD.73.072003;%%

\bibitem{Grange:2015fou} 
Muon g-2 Collaboration: J.~Grange {\it et al.},
Muon $g-2$ Technical Design Report,
arXiv:1501.06858 [physics.ins-det].
%%CITATION = ARXIV:1501.06858;%%

\bibitem{JPARC2011}
M.~Aoki {\it et al.},
Conceptual Design Report for The Measurement of the Muon Anomalous Magnetic Moment $g-2$ and Electric Dipole Moment at J-PARC,\\
\url{https://g2sakura.kek.jp/public/doc/MCDR-submit.pdf}.

\bibitem{Erler:2000nx} 
J.~Erler and M.~Luo,
Hadronic loop corrections to the muon anomalous magnetic moment,
{\em Phys.\ Rev.\ Lett.}\ {\bf 87}, 071804 (2001),
hep-ph/0101010.
%%CITATION = doi:10.1103/PhysRevLett.87.071804;%%

\bibitem{Rodrigo:2001kf} 
G.~Rodrigo, H.~Czyz, J.~H.~K\"uhn and M.~Szopa,
Radiative return at NLO and the measurement of the hadronic cross-section in electron positron annihilation, \\
{\em Eur.\ Phys.\ J.\ C} {\bf 24}, 71 (2002),
hep-ph/0112184.
%%CITATION = doi:10.1007/s100520200912;%%

\bibitem{Davier:2009ag} 
M.~Davier {\it et al.},
The Discrepancy Between $\tau$ and $e^+ e^-$ Spectral Functions Revisited and the Consequences for the Muon Magnetic Anomaly, \\
{\em Eur.\ Phys.\ J.\ C} {\bf 66}, 127 (2010),
arXiv:0906.5443 [hep-ph].
%%CITATION = doi:10.1140/epjc/s10052-009-1219-4;%%

\bibitem{Jegerlehner:2011ti} 
F.~Jegerlehner and R.~Szafron,
$\rho^0$-$\gamma$ mixing in the neutral channel pion form factor $F_{\pi}^{e}$ 
and its role in comparing $e^+ e^-$ with $\tau$ spectral functions, \\
{\em Eur.\ Phys.\ J.\ C} {\bf 71}, 1632 (2011),
arXiv:1101.2872 [hep-ph].
%%CITATION = doi:10.1140/epjc/s10052-011-1632-3;%%

\bibitem{Calame:2015fva} 
C.~M.~Carloni Calame, M.~Passera, L.~Trentadue and G.~Venanzoni, \\
A new approach to evaluate the leading hadronic corrections to the muon $g-2$, \\
{\em Phys.\ Lett.\ B} {\bf 746}, 325 (2015),
arXiv:1504.02228 [hep-ph].
%%CITATION = doi:10.1016/j.physletb.2015.05.020;%%    
  
\bibitem{Abbiendi:2016xup} 
G.~Abbiendi {\it et al.},
Measuring the leading hadronic contribution to the muon $g-2$ via $\mu e$ scattering,
{\em Eur.\ Phys.\ J.\ C} {\bf 77}, 139 (2017),
arXiv:1609.08987 [hep-ex].
%%CITATION = doi:10.1140/epjc/s10052-017-4633-z;%%  
  
\bibitem{Lautrup:1971jf} 
B.~E.~Lautrup, A.~Peterman and E.~de Rafael,
Recent developments in the comparison between theory and experiments in quantum electrodynamics, \\
{\em Phys.\ Rept.}\ {\bf 3}, 193 (1972).
%%CITATION = doi:10.1016/0370-1573(72)90011-7;%%  

\bibitem{Blum:2016lnc} 
T.~Blum {\it et al.},
Connected and Leading Disconnected Hadronic Light-by-Light Contribution to the Muon Anomalous Magnetic Moment with a Physical Pion Mass, \\
{\em Phys.\ Rev.\ Lett.}\ {\bf 118}, 022005 (2017),
arXiv:1610.04603 [hep-lat].
%%CITATION = doi:10.1103/PhysRevLett.118.022005;%%
  
\bibitem{Blum:2017cer} 
T.~Blum {\it et al.},
Using infinite volume, continuum QED and lattice QCD for the hadronic light-by-light contribution to the muon anomalous magnetic moment, \\
{\em Phys.\ Rev.\ D} {\bf 96}, no. 3, 034515 (2017),
arXiv:1705.01067 [hep-lat].
%%CITATION = doi:10.1103/PhysRevD.96.034515;%%  
  
\bibitem{DellaMorte:2017dyu} 
M.~Della Morte {\it et al.},
The hadronic vacuum polarization contribution to the muon $g-2$ from lattice QCD,
{\em JHEP} {\bf 1710}, 020 (2017),
arXiv:1705.01775 [hep-lat].
%%CITATION = doi:10.1007/JHEP10(2017)020;%%

\bibitem{Peskin:1991sw} 
M.~E.~Peskin and T.~Takeuchi,
Estimation of oblique electroweak corrections, \\
{\em Phys.\ Rev.\ D} {\bf 46}, 381 (1992).
%%CITATION = doi:10.1103/PhysRevD.46.381;%%
  
\bibitem{Veltman:1977kh} 
M.~J.~G.~Veltman,
Limit on Mass Differences in the Weinberg Model, \\
{\em Nucl.\ Phys.\ B} {\bf 123}, 89 (1977).
%%CITATION = doi:10.1016/0550-3213(77)90342-X;%%

\bibitem{Erler:2010sk} 
J.~Erler and P.~Langacker,
Precision Constraints on Extra Fermion Generations, \\
{\em Phys.\ Rev.\ Lett.}\ {\bf 105}, 031801 (2010),
arXiv:1003.3211 [hep-ph].
%%CITATION = doi:10.1103/PhysRevLett.105.031801;%%

\bibitem{Erler:2013xha} 
J.~Erler and S.~Su,
The Weak Neutral Current, \\
{\em Prog.\ Part.\ Nucl.\ Phys.}\ {\bf 71}, 119 (2013),
arXiv:1303.5522 [hep-ph].
%%CITATION = doi:10.1016/j.ppnp.2013.03.004;%%

\bibitem{Erler:2014fqa} 
J.~Erler, C.~J.~Horowitz, S.~Mantry and P.~A.~Souder,
Weak Polarized Electron Scattering,
{\em Ann.\ Rev.\ Nucl.\ Part.\ Sci.}\ {\bf 64}, 269 (2014),
arXiv:1401.6199 [hep-ph].
%%CITATION = doi:10.1146/annurev-nucl-102313-025520;%%  

\bibitem{Kumar:2013yoa} 
K.~S.~Kumar, S.~Mantry, W.~J.~Marciano and P.~A.~Souder,
Low Energy Measurements of the Weak Mixing Angle, \\
{\em Ann.\ Rev.\ Nucl.\ Part.\ Sci.}\ {\bf 63}, 237 (2013),
arXiv:1302.6263 [hep-ex].
%%CITATION = doi:10.1146/annurev-nucl-102212-170556;%%

\end{thebibliography}

\end{document}